\documentclass[letterpaper,twocolumn,10pt]{article}
\usepackage{usenix2022_SOUPS}

\usepackage{tikz}
\usepackage{amsmath}

\usepackage{xcolor}
\usepackage{setspace}
\usepackage{multirow}
\usepackage{bm,bbm}
\usepackage{paralist}
\usepackage{graphics}
\usepackage{graphicx}

\begin{document}

\date{}

\title{\Large \bf Can Causal (and Counterfactual) Reasoning improve Privacy Threat Modelling?}

\def\plainauthor{Author name(s) for PDF metadata. Don't forget to anonymize for submission!}

\author{
{\rm Rakshit Naidu}\\
Carnegie Mellon University\\
rnemakal@andrew.cmu.edu
\and
{\rm Navid Kagalwalla}\\
Carnegie Mellon University\\
nkagalwa@andrew.cmu.edu
} 

\maketitle
\thecopyright

\begin{abstract}
Causal questions often permeate in our day-to-day activities. With causal reasoning and counterfactual intuition, privacy threats can not only be alleviated but also prevented. In this paper, we discuss what is causal and counterfactual reasoning and how this can be applied in the field of privacy threat modelling (PTM). We believe that the future of PTM relies on how we can causally and counterfactually imagine cybersecurity threats and incidents.
\end{abstract}

\section{Introduction}

\subsection{Causal reasoning}

Causality originated from the field of psychology and later, pervaded through the field of Computer Science~\cite{pearl2009, Scholkopf2021TowardCR}. Causality refers to how actions are influenced by a cause-and-effect strategy~\cite{matutehe, pearl2009}. For example, if a ball rolls on the floor, the cause could be a person either kicking it off his/her foot or throwing it off his/her hand. This is calculated by probabilities or likelihoods of whether these past events are the causes of the observed event, that is here, the rolling motion of the ball on the floor. ``Correlation does not imply causation''-- a quote that still has significant implications even today. This means that there could be additional variables that potentially influence an outcome. Two variables do not necessarily cause each other to occur and vice versa. A typical example for this effect can be observed in a graph where the ice cream sales and the number of shark attacks have similar trends thereby being highly correlated, given here~\footnote{\href{https://www.statology.org/correlation-does-not-imply-causation-examples/}{Correlation does not imply causation.}} However, it is a well-known fact that consuming ice creams do not have any influence over shark attacks, let alone cause them. Therefore, such reasoning in understanding privacy threats (for example: on how data could be leaked by different potential sources or on how much likelihood does one factor have over another while quantitatively analyzing privacy risks associated to a vulnerability) is crucial. Through equivalence graphs and decision trees, one can causally reason on how various sources could influence cybersecurity threats. In this paper, we specifically follow the definitions of causality as illustrated in~\cite{pearl2009}.

\textbf{Definition 1} \textit{Causal reasoning}

Given a set of known variables $X$ = $\{X_1, ... X_i ..., X_n \}$, a set of outcomes $O$ = $\{O_1, ... O_i ..., O_n\}$, we say that a variable $X_i$ causally affects $O_i$ if there is a direct edge (or relationship) from $X_i$ to $O_i$. 

\subsection{Counterfactual reasoning}

Counterfactual reasoning refers to how one can explain decisions alternatively i.e. in a counterfactual world~\cite{Kusner2017CounterfactualF, Karimi2021AlgorithmicRF, Born2021TheLE}. With counterfactual intuition, we can model and think for cases that have not been observed, which could securely airtight the privacy threat model in question. 

\textbf{Definition 2} \textit{Counterfactual reasoning}

Given a set of known variables $X$ = $\{X_1, ... X_i ..., X_n \}$, a set of unknown variables $X'$ = $\{X'_1, ... X'_i ..., X'_n \}$, a set of outcomes $O$ = $\{O_1, ... O_i ..., O_n\}$, we say that the unknown variable $X'_i$ is a counterfactual variable (could be a confounding variable or intervention) which denotes a counterfactual case $X'_i$ (for the known, factual case $X_i$) which could influence the outcome $O_i$.

Causal and counterfactual reasoning concepts are quite relevant to privacy threat modelling as there are always unseen cases and variables which could lead to privacy disasters, if there are no countermeasures set up. 

\subsection{Relevance of reasoning in the context of privacy threats}

There have been a few papers that explicitly look at causal reasoning and cybersecurity threats~\cite{Abel2020, tople2020alleviating, osti_10296947}. Threat modelling allows companies and organizations to identify potential vulnerabilities in their softwares/systems. Recognizing these flaws enables them to design a framework against such threats.

In this paper, we specifically look at two important (fictitious) threat examples that could be modeled through causal reasoning, as also discussed in~\cite{Abel2020}.

\begin{figure}[t]
    \centering
        
     \includegraphics[width=0.25\textwidth]{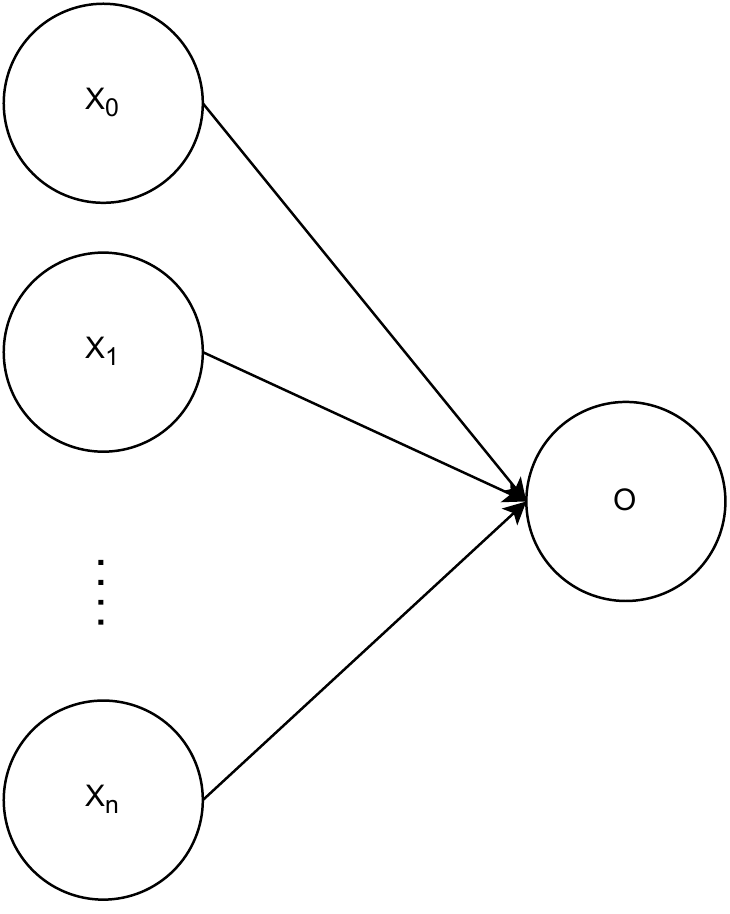}
     \caption{Causal reasoning with various input variables $X_i$ and output variable $O$.}
     \label{fig:fig1}

     \includegraphics[width=0.25\textwidth]{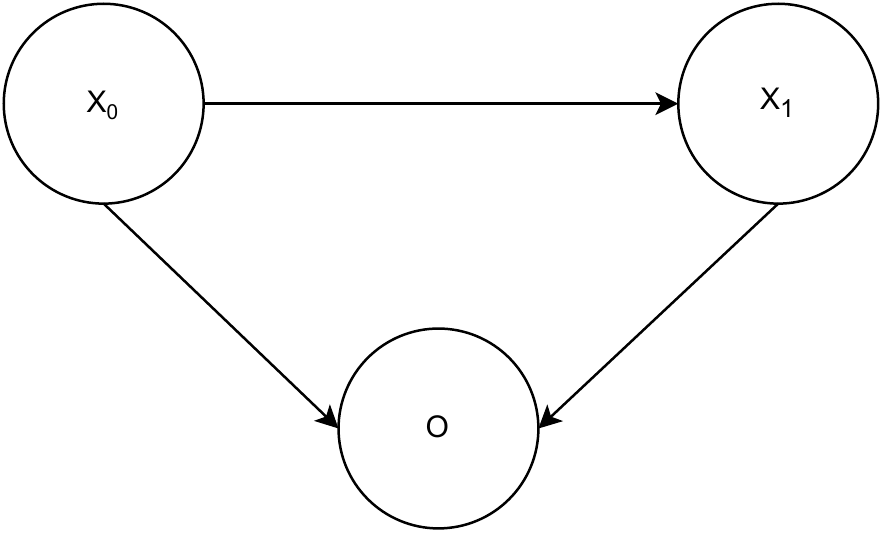}
     \caption{Causal reasoning with confounding variable $X_0$.}
     \label{fig:fig2}
    
    \vspace{-2ex}
\end{figure}

Assuming that the victim's password for his email has been leaked in a data breach of a website (which is the outcome $O$), we can model the different causes ($X$ and $X'$) that could have had effect on why this data breach had led to the victim losing his private and sensitive information, that is, the password. We model this using causal graphs as shown in Figure~\ref{fig:fig1}. 

In Figure~\ref{fig:fig1}, we show cause and effect relation between the input variables to the output. According to our example, the output is the exposure of the victim's password. The input variables could be modelled as how many times the victim may have visited that website ($X_0$), the time spent on that website ($X_1$), hyperlinks clicked on that website ($X_2$) and so on, other input factors that could have led to the victim's password being revealed in a data breach. 
By analyzing the probabilities of the independent input variables given the output (that is, $\Pr[X_{i} | O]$), we can calculate the probability of that outcome $O$ as 

\begin{equation*}
\Pr[O] = \prod^{n}_{0} \left(\Pr[X_{i} | O]\right)
\end{equation*}

By calculating $\Pr[O]$, we could have possibly averted the victim being hacked off his password. 

In Figure~\ref{fig:fig2}, we again show that there could be a confounding variable, which could affect both the input variable as well as the output. 
The number of clicks on a website ($X_0$) could determine how much time he stays on that website ($X_1$) which could mean that his online, browsing activities are more observed by the malicious hacker, which leads to the victim leaking the password ($O$). Counterfactual reasoning could be used to understand other factors as to why such data was breached and whether the victim actually played a pivotal role in the incident. Maybe it could have been malware in the system that led to the incident. Or maybe it was a human error where he accidentally might have typed in the password with an observer watching over him (``shoulder surfing'').

Privacy threat modeling can also be useful in determining an organization’s budget for privacy operations. In this risk-based approach to budgeting, the organization
hopes to arrive at a budget that can reduce, if not mitigate, privacy incidents
completely (the outcome $O$) by modeling the different causes ($X$ and $X'$) that can lead to privacy incidents through assigning probabilities to these causes. The input variables (causes) could be modeled as a violation of compliance requirements set forth by GDPR or CPRA ($X_0$), sensitive data breach due to insecure organizational policies
($X_1$), and unauthorized access to personally identifiable data by third parties ($X_2$) and so on. We can further assign probabilities to these independent input variables depending on the number of times these events occurred in the past or by assessing how competitors view these causes. In this scenario, we can encounter
confounding variables as well, sensitive data breaches due to insecure
organizational policies could lead to a violation of compliance requirements which could impact the mitigation of privacy incidents. Counterfactual
reasoning could also be employed here to assess other factors that could influence
the outcome $O$ such as the size and experience of the privacy team, and the
integration of third-party software that contains vulnerabilities. These would
impact the security posture of the organization which would, in turn, affect the outcome $O$.

\section{Conclusion and Future Work}

In this paper, we provide an overview of causal reasoning and counterfactual reasoning. We also look at an imaginary privacy threat/incident and how we could causally reason about such reports, without it actually occurring. We envision that privacy threat modelling can further be enhanced with causal and counterfactual modelling and thereby, benefit from concepts originally introduced in the field of psychology.

\bibliographystyle{plain}
\bibliography{lib}

\begin{thebibliography}{1}

\bibitem{Abel2020}
Suchitra Abel, Yenchih Tang, Jake Singh, and Ethan Paek.
\newblock {Applications of Causal Modeling in Cybersecurity : An Exploratory
  Approach}.
\newblock {\em Advances in Science, Technology and Engineering Systems
  Journal}, 5(3):380--387, 2020.

\bibitem{Born2021TheLE}
Benjamin Born, Alexander~M. Dietrich, and Gernot M{\"u}ller.
\newblock The lockdown effect: A counterfactual for sweden.
\newblock {\em PLoS ONE}, 16, 2021.

\bibitem{osti_10296947}
A~Ibrahim, S~Rehwald, A~Scemama, F~Andres, and A~Pretschner.
\newblock Causal model extraction from attack trees to attribute malicious
  insider attacks.
\newblock {\em Lecture notes in computer science}, 12419, 2020.

\bibitem{Karimi2021AlgorithmicRF}
Amir-Hossein Karimi, Bernhard Sch{\"o}lkopf, and Isabel Valera.
\newblock Algorithmic recourse: from counterfactual explanations to
  interventions.
\newblock {\em Proceedings of the 2021 ACM Conference on Fairness,
  Accountability, and Transparency}, 2021.

\bibitem{Kusner2017CounterfactualF}
Matt~J. Kusner, Joshua~R. Loftus, Chris Russell, and Ricardo Silva.
\newblock Counterfactual fairness.
\newblock In {\em NIPS}, 2017.

\bibitem{matutehe}
Helena Matute, Fernando Blanco, Ion Yarritu, Marcos Díaz-Lago, Miguel~A.
  Vadillo, and Itxaso Barberia.
\newblock Illusions of causality: how they bias our everyday thinking and how
  they could be reduced.
\newblock {\em Frontiers in Psychology}, 6, 2015.

\bibitem{pearl2009}
Judea Pearl.
\newblock {\em Causality: Models, Reasoning and Inference}.
\newblock Cambridge University Press, USA, 2nd edition, 2009.

\bibitem{Scholkopf2021TowardCR}
Bernhard Scholkopf, Francesco Locatello, Stefan Bauer, Nan~Rosemary Ke, Nal
  Kalchbrenner, Anirudh Goyal, and Yoshua Bengio.
\newblock Toward causal representation learning.
\newblock {\em Proceedings of the IEEE}, 109:612--634, 2021.

\bibitem{tople2020alleviating}
Shruti Tople, Amit Sharma, and Aditya Nori.
\newblock Alleviating privacy attacks via causal learning.
\newblock In {\em International Conference on Machine Learning (ICML)}, July
  2020.

\end{thebibliography}

\end{document}